\documentclass[10pt,draft]{dis03}
\usepackage{epsfig,amsmath}

\newcommand{\LambdaQCD}{\Lambda_{\mbox{\footnotesize QCD}}}

\begin{document}

\title{Some remarks on geometric scaling%
\thanks{This work is funded by the
European Commission IHP program under contract HPRN-CT-2000-00130.}}

\author{St\'ephane Munier\footnote{Permanent address: Centre de physique th\'eorique, 
\'Ecole polytechnique, 91128 Palaiseau cedex, France. 
Email: Stephane.Munier@cpht.polytechnique.fr} \\
Institute for theoretical physics, University of Heidelberg\\
Philosophenweg 19,
69120 Heidelberg, Germany.\\
E-mail: munier@tphys.uni-heidelberg.de }

\maketitle

\begin{abstract}
\noindent 
In order to explain a striking symmetry of the HERA data for virtual
photon-proton cross section,
I propose a simple model based on the
elementary 2-gluon exchange dipole-dipole cross section, 
and which exhibits geometric scaling.
I also suggest that geometric scaling should manifest itself in
exclusive processes. A preliminary search for this property in the HERA
data is presented. 
\end{abstract}

\section{Geometric scaling}

Geometric scaling \cite{Stasto:2000er} refers to the property that the
total virtual photon-proton
cross section $\sigma(x,Q)$ at small-$x$ depends on the combined variable
$Q/Q_s(x)$ only, where
\begin{equation}
Q_s(x)=1\ \mbox{GeV}\cdot\left({x}/{x_0}\right)^{-\lambda_0/2},\ 
\mbox{with}\ x_0\sim3\times 10^{-4}\ \mbox{and}\ \lambda\sim 0.3
\label{eq:qs}
\end{equation} 
is the so-called saturation scale. This scale is related to the mean
transverse momentum of the partons inside the proton. Its
growth with $1/x$ stems from gluon recombination 
(for a review, see \cite{Mueller:2001fv}).
The fact that geometric scaling holds 
is often considered a hint for saturation
effects in deep-inelastic scattering.

Our first remark concerns
a point that has not been much commented so far: the almost perfect
symmetry of $Q/Q_s\times\sigma(Q/Q_s)$ under the interchange of $Q$
and $Q_s$~\cite{Stasto:2000er}. We point out that this symmetry 
is already present in the 2-gluon exchange dipole-dipole cross section
and starting from this observation, we propose a simple model for $\sigma$.
Our second remark is that 
the saturation scale should be understood as an impact
parameter-dependent quantity $Q_s(x,b)$, and that exclusive amplitudes
exhibit {\it local} geometric scaling, in the sense that they should
depend on $Q/Q_s(x,b)$ only.

\section{A simple model}

The elementary dipole-dipole cross section of two dipoles of respective
size $1/Q$ and $1/Q_s$ reads
\begin{equation}
\sigma_{dd}(1/Q,1/Q_s)=2\pi\alpha_s^2(\min(1/Q,1/Q_s))^2
\left(1+\left|\log({Q}/{Q_s})\right|\right)\ .
\end{equation}
The fundamental observation is that the magnitude of the cross section
is given by the squared size of the {\it smallest} interacting dipole.

Let's try a simple model for $\sigma$ \cite{Munier:2002gf}. Assume that
the proton is a collection of {\it independent} dipoles of size
$1/Q_s$ uniformly
spread over a surface $1/\LambdaQCD^2$. The photon is itself
represented by $q\bar q$ dipoles of
size $1/Q$ whose flux we approximate by a constant $N$. 
The interaction proceeds through the
exchange of two gluons between the photon dipole and one of the
dipoles in the proton. With these simple assumptions, 
the $\gamma^*p$ cross section reads
\begin{equation}
\begin{split}
\sigma&=\left\{\mbox{flux of dipoles in the photon}\right\}\\
 &\phantom{flux of}\times\left\{\mbox{number of dipoles in the proton}
\right\}\times \sigma_{dd}\\
&=N \frac{1}{\LambdaQCD^2}2\pi\alpha_s^2
 \min(Q_s^2/Q^2,1)(1+|\log(Q/Q_s)|)\ .
\end{split}
\end{equation}
It is easy to see that $Q/Q_s\times\sigma(Q/Q_s)$
is symmetric under the replacement $Q\leftrightarrow Q_s$.
For $Q\gg Q_s$, this model is equivalent to
the BFKL evolution with appropriate boundary conditions, 
where the QCD dynamics determine the energy evolution of $Q_s(x)$.
For $Q\ll Q_s$, it turns out that this model can be qualitatively understood as an
effective picture for color glass condensate~\cite{Iancu:2002aq}.

However, this model has to be supplemented by multiple gluon-pair exchanges
when $Q\sim Q_s$, which we implemented in a Glauber-Mueller inspired way. 
Indeed, in this kinematical region, the probability for dipole-dipole
interaction can become quite large even for moderate values of
$\alpha_s$.
The resulting model, depending on 5 free parameters, 
gives a good fit to all recent ZEUS
data~\cite{Chekanov:2001qu} for $F_2$
($\chi^2=1.15/\mbox{d.o.f}$), 
as can be seen on Fig.1 (for details, see~\cite{Munier:2002gf}).

\begin{figure}[!thb]
\vspace*{10.5cm}
\begin{center}
\includegraphics{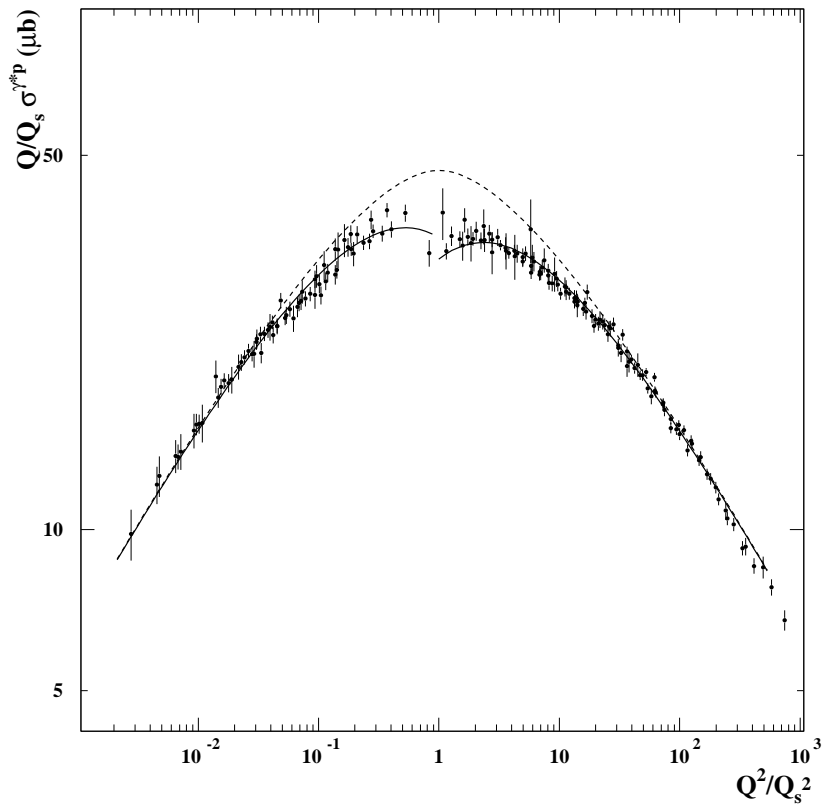}
\caption{An illustration of geometric scaling and the symmetric
  saturation model. Data from Ref.\cite{Chekanov:2001qu}}
\end{center}
\end{figure}

The success of the obtained parametrization is also a first step toward
the quantitative estimate of the statistical relevance of geometric scaling.

\section{Local geometric scaling}

So far, 
most theoretical and phenomenological discussions on saturation have assumed
a unique saturation scale $Q_s(x)$. However, the value of the saturation scale
is related to the quark and gluon density, thus 
for a realistic proton, 
it is dependent on the impact parameter~$b$. A deeper
understanding of saturation effects thus requires to take into account the
transverse profile of the nucleus. Some progress has been made very
recently on the
theoretical side (see for example
\cite{Ferreiro:2002kv,Golec-Biernat:2003ym}, and
{\cite{Kowalski:2003hm}} for a phenomenological model), and as pointed out in
Ref.\cite{Munier:2001nr}, it is possible to extract the $b$-profile
from experimental data.

We have shown in Ref.\cite{Munier:2003bf} that amplitudes 
${\cal  A}(x,Q,b)$ which are
explicitely dependent on~$b$, like elastic diffraction (in practice:
diffractive production of vector mesons of mass $M$) exhibit geometric scaling, i.e.
\begin{equation}
{\cal A}(x,Q,b)={\cal A}((Q^2\!+\!M^2)/Q_s^2(x,b))\ .
\end{equation}
We have checked that this statement, which can be derived from QCD, is
supported by experimental measurements.
On Fig.2, we represent the amplitude $\cal A$ that we deduced from the
present data on elastic $\rho$ meson
production, as a function of the scaling variable $(Q^2\!+\!M^2)/Q_s^2(x,b)$, for 3
different values of $b$ (for each $b$, the form~(\ref{eq:qs}) was
assumed for $Q_s(x,b)$). One sees that the result is a good hint for
local geometric scaling.

\begin{figure}[!thb]
\vspace*{10.5cm}
\begin{center}
\includegraphics{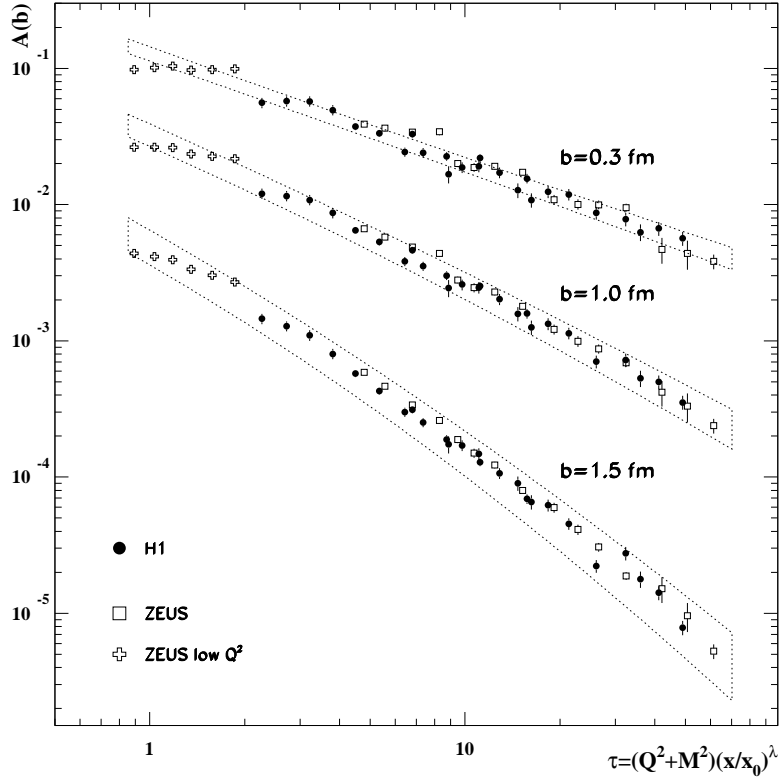}
\caption{Local geometric scaling. The data were taken from
  Ref.\cite{Adloff:1999kg}. Note however that the
  displayed points are not true measurements, but are
only {\it derived} from experimental data through
  the procedure explained in~Ref.\cite{Munier:2003bf}. The bands give
  an (upper) estimate of the uncertainties.}
\end{center}
\end{figure}

\section{Conclusion and outlook}

Geometric scaling may be considered as an indirect manifestation of
saturation effects, and
has been understood within saturation models. In
this talk, we have essentially
advocated the study of local geometric scaling, which would give
deeper insight into saturation physics.
It requires the experimental extraction of
scattering amplitudes at fixed impact parameter, which we believe is
doable with the present experimental data on diffraction.


\begin{thebibliography}{0}


\bibitem{Stasto:2000er}
A.~M.~Sta\'sto, K.~Golec-Biernat and J.~Kwieci\'nski,
Phys.\ Rev.\ Lett.\  {\bf 86} (2001) 596
[arXiv:hep-ph/0007192].

\bibitem{Mueller:2001fv}
A.~H.~Mueller,
arXiv:hep-ph/0111244, and summary talk at this conference.

\bibitem{Munier:2002gf}
S.~Munier,
Phys.\ Rev.\ D {\bf 66} (2002) 114012
[arXiv:hep-ph/0205319].

\bibitem{Iancu:2002aq}
E.~Iancu, K.~Itakura and L.~McLerran,
Nucl.\ Phys.\ A {\bf 724} (2003) 181
[arXiv:hep-ph/0212123].

\bibitem{Chekanov:2001qu}
S.~Chekanov {\it et al.}  [ZEUS Collaboration],
Eur.\ Phys.\ J.\ C {\bf 21} (2001) 443
[arXiv:hep-ex/0105090];
J.~Breitweg {\it et al.}  [ZEUS Collaboration],
Phys.\ Lett.\ B {\bf 487} (2000) 53
[arXiv:hep-ex/0005018].


\bibitem{Ferreiro:2002kv}
E.~Ferreiro, E.~Iancu, K.~Itakura and L.~McLerran,
Nucl.\ Phys.\ A {\bf 710} (2002) 373
[arXiv:hep-ph/0206241].

\bibitem{Golec-Biernat:2003ym}
K.~Golec-Biernat and A.~M.~Sta\'sto,
arXiv:hep-ph/0306279, and talk by A.~M.~Sta\'sto at this conference.

\bibitem{Kowalski:2003hm}
H.~Kowalski and D.~Teaney,
arXiv:hep-ph/0304189, and talk by H.~Kowalski at this conference.

\bibitem{Munier:2001nr}
S.~Munier, A.~M.~Sta\'sto and A.~H.~Mueller,
Nucl.\ Phys.\ B {\bf 603} (2001) 427
[arXiv:hep-ph/0102291].

\bibitem{Munier:2003bf}
S.~Munier and S.~Wallon, to appear in Eur. Phys. J. C (2003)
[arXiv:hep-ph/0303211].

\bibitem{Adloff:1999kg}
C.~Adloff {\it et al.}  [H1 Collaboration],
Eur.\ Phys.\ J.\ C {\bf 13} (2000) 371
[arXiv:hep-ex/9902019];
J.~Breitweg {\it et al.}  [ZEUS Collaboration],
Eur.\ Phys.\ J.\ C {\bf 6} (1999) 603
[arXiv:hep-ex/9808020].



\end{thebibliography}
\end{document}